# Signature of geometry modulation on interface magnetism emerged in isomeric $IrO_2$-$CoFe_2O_4$ heterostructures

Meng Wang[1,*], Shunsuke Mori[1], Xiuzhen Yu[1], Masahiro Sawada[2], Naoya Kanazawa[3], Pu Yu[4], Fumitaka Kagawa[1,5]

[1]RIKEN Center for Emergent Matter Science (CEMS), Wako, 351-0198, Japan.

[2]Hiroshima Synchrotron Radiation Center, Hiroshima University, Higashihiroshima, Hiroshima 739-0046, Japan.

[3]Institute of Industrial Science, The University of Tokyo, Tokyo 153-8505, Japan.

[4]State Key Laboratory of Low Dimensional Quantum Physics and Department of Physics, Tsinghua University, Beijing 100084, China.

[5]Department of Physics, Tokyo Institute of Technology, Tokyo 152-8551, Japan

*Corresponding author. Email: meng.wang@riken.jp;

**The interface composed of magnets and strong spin-orbit coupling (SOC) materials forms an important platform for spintronic devices and intriguing magnetic phenomena, such as the chiral spin textures[1-6] and magnetic proximity effect (MPE)[7-12]. The interface exchange interaction and Dzyaloshinskii–Moriya interaction (DMI) have been discussed in a wide range of heterostructures, while the crystal stacking geometry modulation on these interface interactions has rarely been considered. Here, we show a pronounced geometry modulation on the interface magnetism through comparing a rutile and an anatase $IrO_2$ capping on a ferrimagnetic $CoFe_2O_4$. The rutile heterostructure with a high-symmetry interface shows a conventional anomalous Hall effect (AHE) profile due to the MPE. In contrast, the anatase one with a low-symmetry interface exhibits a topological-like AHE even at zero-field, suggesting the emergence of non-coplanar magnetic order at the interface. Our results suggest that the influence of DMI at the interface can be more accentuated by forming a low-symmetry interface and raises a new means of designing interface magnetism via the geometry modulation.**

When a ferromagnet and a strong SOC paramagnetic-metal (PM) come into contact and form an interface, the combination of the inversion symmetry breaking at the interface and the strong SOC could lead to a finite DMI in the magnetic layer, which furthermore could generate a non-coplanar spin texture，such as skyrmion, as reported in some systems[1-6]. At the same time, through the exchange interaction and/or charge transfer effect across the interface, the magnetic texture near the interface in the ferromagnetic material can have an impact on the PM material near the interface, potentially leading to a spin-ordered state in the PM layer, a phenomenon known as the magnetic proximity effect (as shown in Supplementary Fig. 1) [7-12].

In general, the MPE-induced spin-order in the PM material is confined within a thickness of ~1 nm, because the impact of the interface exchange interaction decays dramatically into the PM material as being away from the interface [12]. Besides, the induced spin order near the interface has often been assumed to be either parallel or antiparallel to the magnetic layer, depending on the signs of the exchange interactions. The interface exchange coupling energy (~meV) is generally larger than Zeeman energy (at ~ 10 T),[13] and hence the MPE-induced spin order within the PM layer exhibits a switching concomitantly with the magnetization reversal of the ferromagnetic layer (Supplementary Fig. 1). Thus, the interface magnetism can be expected to be probed by the AHE measurement of the PM layer.

The interface-induced DMI in the FM material works exclusively in the top layer adjacent to the strong SOC ions.[14] It imparts a handedness to the magnetic exchange interaction with a so-called DM vector and could give rise to the formation of chiral spin textures in some heterostructures [1-6]. Note that the DM vector depends on both the strength of SOC and the bonding geometry. Compared with the interfaces in which the PM and the ferromagnetic layers share a similar crystal symmetry,[3-5] the interface built with materials belonging to different crystal space groups could form a low-symmetry stacking geometry and strongly modulate the interface DMI in some systems. However, such an effect has not yet been explored clearly.

Here, we grew isomeric binary oxides $IrO_2$ on a spinel $CoFe_2O_4$ (CFO) for comparison and studied the interface magnetic states by AHE measurements. The CFO (space group, Fd3m) is a ferrimagnetic insulator due to the antiparallel spin coupling of the octahedral ($O_h$) Fe and tetrahedral ($T_d$) Co sites with a cubic structure (a = 8.38 Å).[15] It can be grown on MgO substrate (cubic, a = 4.21 Å) with a tiny strain. Prior first principle calculations predicted that the $IrO_2$ can form different crystal structures, while the stable phase that has been reported is the rutile (space group, $P4_2/mnm$, a = b = 4.54 Å, c = 3.81 Å)[16,17]. Here, we succeeded in fabricating a metastable phase, anatase (space group, $I4_1/amd$, a = b = 3.92 Å, c = 9.78 Å),[17] by tuning the growth condition of $IrO_2$ films (See Methods).

The structures of the interfaces can be first roughly considered from the lattice-constant point of view. The expected interface structures with rutile and anatase $IrO_2$ are illustrated in **Figs. 1a,b** and **Figs. 1c,d**, respectively. The lattice mismatch and crystal symmetry determine that the rutile $IrO_2$ will grow on CFO along the [1 1 0] as the out-of-plane orientation, while the anatase along the [1 0 0] (See methods). As will be discussed later, this expectation is experimentally confirmed by our X-ray diffraction (XRD) results. The combination of the rutile structure and CFO is expected to maintain two mirror planes orthogonal to the interface as a symmetry operation ($\mathcal{M}$), as illustrated in Fig. 1b. In contrast, for the stacking of the anatase $IrO_2$ on CFO, a much larger lattice mismatch between the anatase [0 0 1] and CFO [1 0 0] inhibit the formation of such a high-symmetry interface, thereby leading to a more complex connection between the Ir and Fe/Co ions across the interface (Figs. 1c, d). As a result, the two mirror planes are lost, and thus, the anatase/CFO has a lower symmetry than the rutile/CFO. Moreover, the distribution of the Ir ions in anatase/CFO interface will potentially exhibit more kinds of crystallographically inequivalent sites than the rutile/CFO, as shown in Supplementary Fig. 2. Inspired by the observations of the enhanced DMI in asymmetrical sandwich structures, such as the Pt/Co/Ta [4] and the $LaO/MnO_2/SrO/IrO_2$,[5] we expect that the reduced symmetry due to the complex connection between the Ir and Fe/Co ions could lead to a larger DMI, thus having more chance to drive the interface magnetism toward a non-coplanar order. If a spin chirality ($\boldsymbol{\Omega}$)[18,19] remains finite in such a non-coplanar spin texture (**Fig. 1e**), one could expect a corresponding topological Hall feature in the transport measurements.

The $IrO_2$/CFO heterostructures with different $IrO_2$ phases were first verified by XRD measurements (See Methods). As shown in **Fig. 2a**, the diffraction peaks at 27.8° and 47.4° are in excellent agreement with the rutile (1 1 0) and anatase (2 0 0), respectively[16,17]. Besides, both heterostructures show a single phase without other XRD diffraction peaks. Compared to the rutile phase, the anatase one was fabricated for the first time in this work, to the best of our knowledge, which was further confirmed by a cross-sectional scanning transmission electron microscopy (STEM) measurement (See Methods). As shown in Supplementary Fig. 3, both the atomic column image and electron diffraction consistently confirm the successful fabrication of the anatase $IrO_2$ on a CFO layer. The energy-dispersive spectrometry (EDS) mappings suggest the ionic diffusion over the interface is negligible, which is attributable to the large difference of ionic radii among Fe, Co and Ir ions, as well as the relatively low growth temperature (~300°C). The CFO surface morphology with a roughness of half unit-cell level further guarantees a high-quality interface on a large scale (Supplementary Fig. 4). In addition, to investigate the valence states at the interfaces, we performed the soft X-ray absorption spectroscopy (XAS) on Fe and Co $L$-edges (See Methods). As shown in **Fig. 2b** and Supplementary Fig. 5, the samples capped with two types of $IrO_2$ show almost identical spectral features in both Fe $L$-edge and Co $L$-edge, indicating nearly identical valence in the two phases. The magnetic state of $IrO_2$ itself was studied by investigating the films directly grown on bare nonmagnetic MgO substrates. We observed the linear magnetic-field-dependent Hall resistivity and small positive magnetoresistance (MR), as shown in the Supplementary Fig. 6, concluding the paramagnetic metallic states of both rutile and anatase $IrO_2$ .

**Figure 2c** shows the temperature-dependent resistivity ($\rho$–$T$) curves of the $IrO_2$ grown on CFO, demonstrating a metallic state down to the lowest temperature in both phases. The anatase shows higher conductivity than the rutile, and this situation is similar to the case of anatase and rutile $TiO_{2-\delta}$, in which a key role of the electron-phonon interaction in the transport properties is discussed.[20,21] To probe the interface magnetism, we measured the Hall resistivity and studied AHE on both $IrO_2$/CFO heterostructures with the Hall bar illustrated in Fig. 2c, inset (see methods). As shown in **Fig. 2d** and Supplementary Fig. 7, the rutile and anatase $IrO_2$ grown on CFO show *p*-type and *n*-type carriers, respectively, consistent with the results on the $IrO_2$ directly grown on the nonmagnetic MgO substrate (Supplementary Fig. 6). To obtain the AHE signal, we decompose the Hall resistivity data into two contributions, i.e., ordinary Hall resistivity proportional to the external magnetic field ($H$) with a coefficient $R_o$, and anomalous Hall resistivity ($\rho_{yx}^{AHE}$) [22]:

$$\rho_{yx} = R_o \cdot H + \rho_{yx}^{AHE}. \qquad (1)$$

Accordingly, by subtracting the ordinary Hall contributions (See Methods), the anomalous Hall resistivity of the two heterostructures can be obtained at various temperatures. **Figures 3a** and **3b** display the $\rho_{yx}^{AHE}$-$H$ curves obtained in two heterostructures, respectively. The magnetic-field-dependent magnetization ($M$-$H$) curves of the CFO films are also shown in gray for comparison.

We first focus on the rutile $IrO_2$ on CFO. Note that in the temperature range of 10–100 K, the $M$-$H$ curves of CFO and the $\rho_{yx}^{AHE}$-$H$ curves of rutile $IrO_2$ invariably exhibit almost identical profiles and coercive field ($H_c$) (Fig. 3a), indicating $\rho_{yx}^{AHE} \propto M$ and a concomitant switching of the MPE-induced magnetic order in rutile $IrO_2$ and the magnetic moment of CFO. The tiny difference in the $M$-$H$ and $\rho_{yx}^{AHE}$-$H$ profiles is seen at low temperatures, but this is attributable to the fact that the maximum magnetic field for the $M$-$H$ measurements (7 T) is slightly smaller than the saturation field of the CFO.

In contrast, we find that the $\rho_{yx}^{AHE}$-$H$ curves in the anatase phase show qualitatively different behaviour from those in the rutile phase (Fig. 3b). While the $\rho_{yx}^{AHE}$-$H$ curves in the anatase phase show the identical sign and profile with the rutile one at (or above) 50 K, the $\rho_{yx}^{AHE}$-$H$ curve with a reversed sign and a reversed direction of hysteresis loop is observed at the lowest temperature; i.e., the coercive magnetic field, $H_c$, is apparently negative (Fig. 3b, 3 K). This unconventional hysteretic behaviour unique to the anatase phase can be seen even in the raw $\rho_{yx}$-$H$ curve at 3 K (Fig. 2d). Assuming that the $\rho_{yx}^{AHE}$-$H$ is saturated at 14 T, both the magnetic moments of anatase $IrO_2$ induced by MPE and the CFO should be aligned toward the direction of the external magnetic field. Such a sign reversal of the $\rho_{yx}^{AHE}$ with decreasing temperature has been reported for the ferromagnetic $SrRuO_3$, and the sign reversal of the momentum-space Berry curvature is thought to underlie the observation.[3,23] A similar mechanism may account for the temperature dependence of $\rho_{yx}^{AHE}$ at the magnetic interface of anatase. The sign reversal $\rho_{yx}^{AHE}$ occurs at the intermediate temperature, 20 K, at which $\rho_{yx}^{AHE}$ (14 T) becomes nearly zero (Fig. 3b). However, a hump-like feature emerges in the $\rho_{yx}^{AHE}$-$H$ profile at low fields, which can not be interpreted by an artifact of the process of subtracting ordinary Hall resistivity (Supplementary Fig. 8), suggesting that an additional term that is not proportional to $M$ is intrinsic and should be considered in more detail to understand the whole profile [3,5,24].

To gain more insight into the $\rho_{yx}^{AHE}$-$H$ profile that is not proportional to magnetization, we further assumed that the anomalous Hall conductance, $G_{xy}^{AHE}$ (= $\rho_{yx}^{AHE} \cdot \rho_{xx}^{-2} \cdot t$; $t$ is the thickness of $IrO_2$), of the anatase heterostructure consists of two contributions: [24]

$$G_{xy}^{AHE}(H) = G_{xy}^{M}(H) + G_{xy}^{T}(H) \qquad (2)$$

$$= S_H M(H) + G_{xy}^{T}(H), \qquad (3)$$

where the $G_{xy}^{M}$ and $G_{xy}^{T}$ denote the anomalous Hall contributions proportional to the uniform magnetization ($M$) and the spin chirality ($\boldsymbol{\Omega}$, as shown in Fig. 1e)[19], respectively. The latter contribution may be referred to as the topological Hall effect [3,5,19,24]. Considering that the $G_{xy}^{M}(H)$ curve should be an identical profile to the $M$-$H$ curve at each temperature in both the rutile and anatase heterostructures, the $G_{xy}^{M}(H)$ curve of the anatase heterostructure is expected to be proportional to the $G_{xy}^{M}(H)$ of the rutile one, which is $\approx G_{xy}^{AHE}(H)$ because of the absence of $G_{xy}^{T}(H)$ for the rutile one. The coefficient, $S_H$, for the anatase can be determined by fitting the $G_{xy}^{AHE}(H)$, where all the magnetic moments are aligned toward the external field (at 14 T, for instance). Thus, the $G_{xy}^{T}(H)$ can be obtained by calculating $G_{xy}^{AHE}(H) - S_H M(H)$. The resulting decomposition of $G_{xy}^{AHE}(H)$ for the case of 3 K is shown in **Fig. 4a**. Distinctly, the resulting $G_{xy}^{T}$ appears at –9 T (+9 T), becomes maximum at +2 T (–2 T), which is close to the magnetization reversal of CFO, and disappears at +6 T (–6 T) in the field increasing (decreasing) process. Thus, the magnetic field dependence of $G_{xy}^{T}$ exhibits a hump feature. The calculated $G_{xy}^{T}$ is summarized as a contour map in the *H*-*T* space (**Fig. 4b**), in which only the data measured during the magnetic field sweeping from -14 T to +14 T are shown for clarity. The contour map clearly shows that $G_{xy}^{T}$ signal decreases gradually as the temperature increases and eventually disappears at ~50 K. Note that the hypothetical two channels mechanism based on the coexisting opposite-sign $G_{xy}^{AHE}(H)$ with non-synchronous switching of the magnetization can also exhibit a hump feature [25,26]. However, this model does not produce negative $H_c$ in the AHE curves and therefore appears irrelevant to the present system. Thus, the emergent $G_{xy}^{T}$ signal suggests that a non-coplanar spin texture with a finite spin chirality feature is formed in the anatase/CFO interface, such as the magnetic skyrmions,[1-6] the textures analogous to that of $CoTa_3S_6$ [27] or $Pr_2Ir_2O_7$ [28], etc [29].

Although the details of the non-coplanar magnetic order in the anatase/CFO heterostructure remain unclear due to the difficulty of direct observation of the interface magnetism, the contrasting behaviour of the AHE in the two isomeric heterostructures indicates that lowering the symmetry of the crystal stacking geometry is an effective way to produce a topological-Hall-like signal in a heterostructure consisting of a paramagnetic metal and an insulating ferromagnet. Note that the isomeric materials are widespread in binary oxides, such as rutile, anatase, λ-phase, etc [20,21,30]. We envision that the concept of isomeric heterostructure can be extended into more systems to design novel electronic and magnetic states.

## Methods:

**Synthesis of $IrO_2$/CFO heterostructures.** The heterostructures were fabricated by pulsed laser deposition method equipped with an excimer laser ($\lambda$ = 248 nm). The CFO films were first grown on MgO (001) substrate with a temperature of 450°C, and oxygen pressure of 15 mTorr, the energy fluence was controlled at 1.5 $J/cm^2$ and the frequency was 5 Hz. The thickness of CFO for transport and magnetization measurements were controlled at ~ 20 nm. After that, the rutile $IrO_2$ was fabricated with a relatively high oxygen pressure of 40 mTorr, a temperature of 350-380 °C, a fluence of 1.3 $J/cm^2$, and a frequency of 2 Hz. The anatase $IrO_2$ phase was obtained with a growth oxygen pressure of 8 mTorr, a temperature of 270~300°C, a fluence of 0.8 $J/cm^2$, and a frequency of 7 Hz. After the depositions, all films were annealed with a high oxygen pressure of 10 Torr at 200°C for 15 minutes to reduce the oxygen vacancy and then cooled to room temperature at a rate of 10 °C/min. The in-plane [-1 1 0] and [0 0 1] of rutile have a mismatch of ~7.2% and ~6.4% to the [1 -1 0] and [1 1 0] of CFO, respectively. The in-plane [0 1 0] of anatase has a lattice mismatch of ~ 6.8% to the [0 1 0] of CFO, while the [0 0 1] of anatase has a huge mismatch of ~16.1% to the [1 0 0] of CFO.

**XRD and XAS measurements**. The XRD $2\theta$-$\omega$ scans were carried out with a high-resolution system (Smartlab, Rigaku) with a Cu-$K_{\alpha 1}$ ($\lambda$ = 0.15406 nm). The XAS curves of Fe and Co were measured at room temperature in BL14, Hiroshima Synchrotron Radiation Center (HiSOR), Hiroshima University with a total electron yield mode. The curves were normalized by subtracting a background.

**STEM measurements**. The sample was prepared by mechanical polishing and Ar-ion milling process. The cross-sectional scanning transmission electron microscopy (STEM) and corresponding element-mapping images were acquired using a TEM (JEM-2800, JEOL) equipped with energy-dispersive X-ray spectrometry (EDS) at an accelerating voltage of 200 kV. These results were analyzed using an X-ray microanalysis system (NORAN SYSTEM 7, Thermo Scientific). The cross-sectional high-resolution transmission electron microscopy (HRTEM) images were acquired using a TEM (JEM-2800, JEOL) at an accelerating voltage of 200 kV.

**Anomalous Hall measurements**. The Hall bars with a size of 300 μm × 60 μm were used for all measurements, which were fabricated by a photo-lithography and an Ar-ions milling system. The transport measurements were conducted with a 14 T PPMS system (Quantum Design) with an in-plane DC current of ~1.0 × $10^5$A /$cm^2$. The Hall resistivity $\rho_{yx}$ was measured by $\rho_{yx} = V_{yx} / I_{xx} \cdot t$, where $V_{yx}$, $I_{xx}$, and $t$ correspond to the transverse voltage, longitudinal current, and the $IrO_2$ film thickness, respectively. The longitudinal magnetoresistance was subtracted by the method in the reference[31]. The anomalous Hall resistivity $\rho_{yx}^{AHE}$ was obtained by subtracting the ordinary Hall contribution, which was determined with a linear fitting on the high magnetic field regions. The anomalous Hall conductance $G_{xy}^{AHE}$ was obtained by $G_{xy}^{AHE} = \rho_{yx}^{AHE} / \rho_{xx}^2 \cdot t$.

**Acknowledgments:** This study was financially supported by JSPS KAKENHI (Grant Nos. 21H04442, 21K14398) and JST CREST (Grant No. JPMJCR1874). P. Y. acknowledges the supports from the National Key R&D Program of China (grant No. 2021YFE0107900) and the National Natural Science Foundation of China (grant No. 52025024).

**Data availability:** All the data presented in the article and Supplementary Information are available from the corresponding authors upon reasonable request.

**Author contributions:** M. W. and F. K. conceived the project. M. W. fabricated the samples with the help of P. Y. M. W. performed the magnetization, XRD, and transport measurements. S. M. and X. Y. performed the STEM measurements. M. W. performed the XAS measurements with the help of M. S. N. K. and P. Y. provided deep discussions. M. W. and F. K. wrote the manuscript. All authors discussed the results and commented on the manuscript.

**Competing Interests:** The authors declare no competing interests.

## Figures

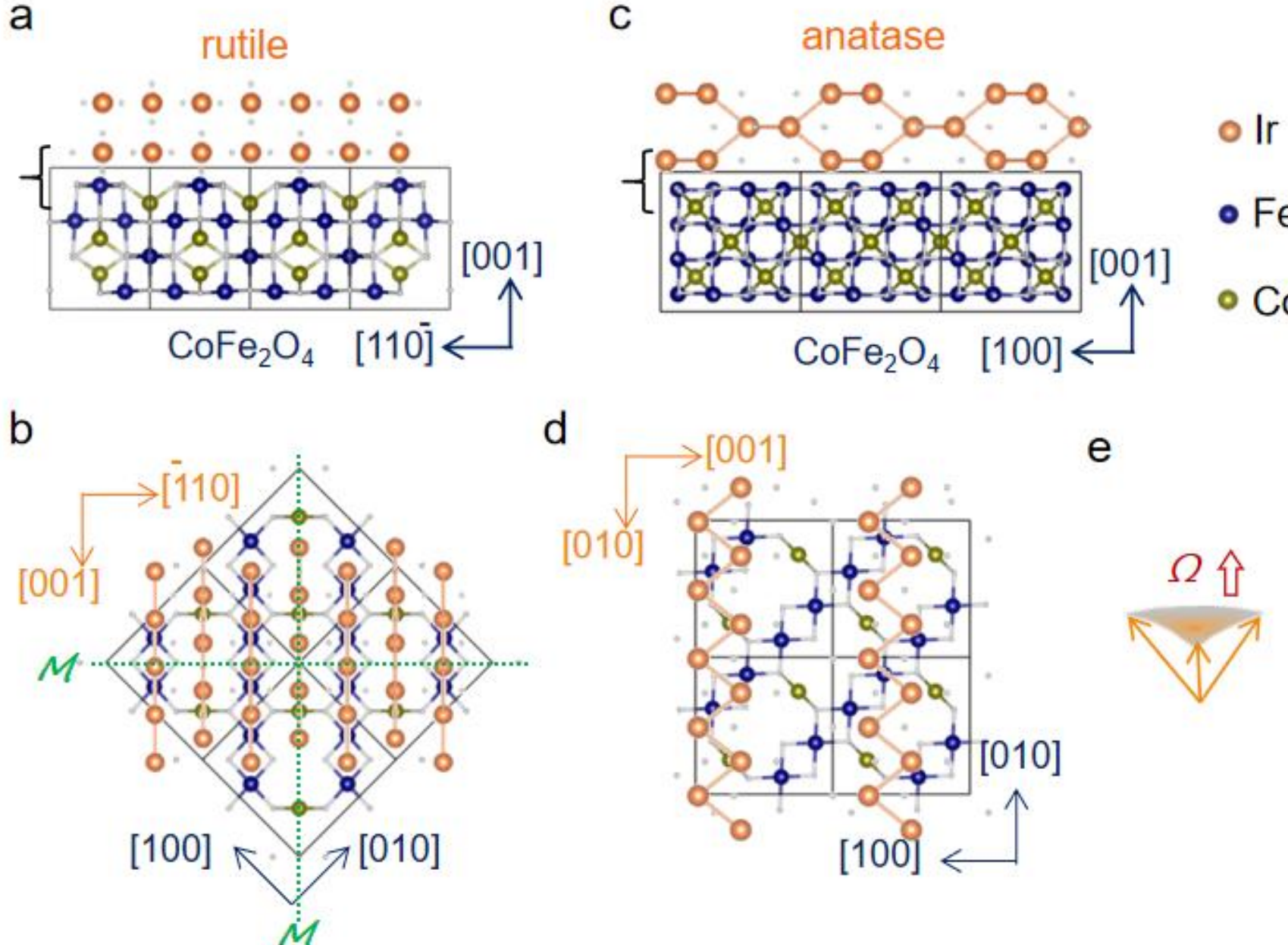

**Figure 1. Illustration of the atomic stacking and magnetism at two kinds of interfaces. a**, **c**, Crystal structures of the $IrO_2/CoFe_2O_4$ (CFO) interfaces with a side-view for rutile (a) and anatase (c) $IrO_2$, respectively. **b**, **d**, Top-views of the $IrO_2$/CFO interfaces with the rutile (c) and anatase (d) capping on CFO, respectively. The crystal orientations of CFO and $IrO_2$ are determined by minimizing the lattice mismatch, and are shown by the blue and orange axis, respectively. The mirrors ($\mathcal{M}$) of the of the rutile type interface are shown by green lines in (b). The oxygen ions are shown with small gray spheres. **e**, Illustration of the spin chirality ($\boldsymbol{\Omega}$) due to the non-coplanar magnetic order.

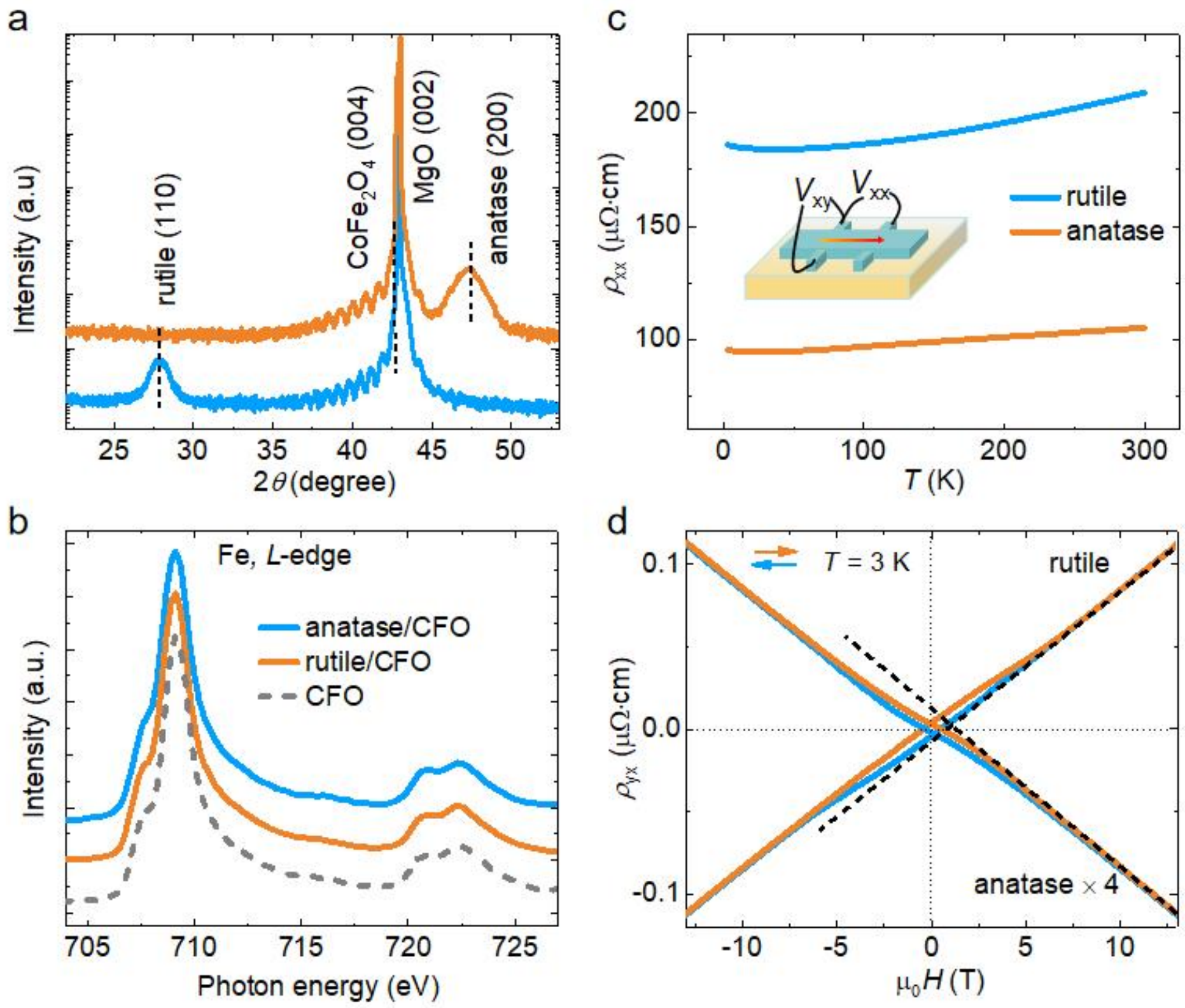

**Figure 2. Characteristics of the two kinds of heterostructures. a**, XRD $2\theta$-$\omega$ scans of the two heterostructures built with CFO and rutile or anatase $IrO_2$, respectively. **b**, XAS around Fe $L$-edges measured on two heterostructures, with a comparison to that from bare CFO. Capping layers of anatase and rutile $IrO_2$ are 2 nm. **c**, Temperature dependent resistivity ($\rho$–$T$) curves of the rutile and anatase $IrO_2$ films grown on CFO. Inset, an illustration of the Hall bar for the transport measurement. **d**, Raw data of the Hall resistivity at 3 K measured from the two heterostructures. Dash lines indicate the ordinary Hall contributions with a linear fitting at high magnetic field regions. The thickness of $IrO_2$ is 5 nm in rutile and 4 nm in anatase.

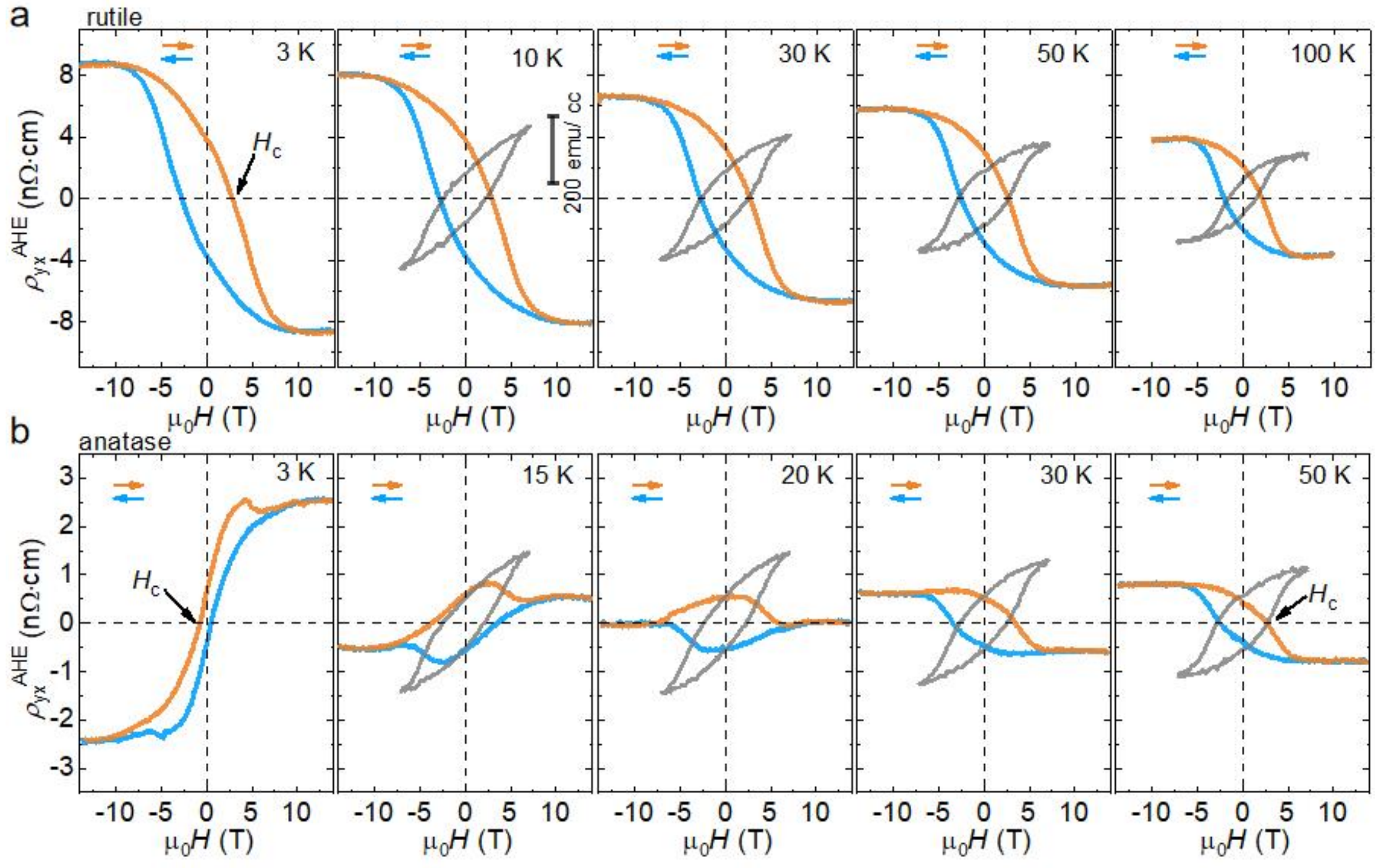

**Figure 3. Comparing AHE curves of the two kinds of $IrO_2$/CFO heterostructures. a**, **b**, Magnetic field dependent anomalous Hall resistivity ($\rho_{yx}^{AHE}$-$H$) curves at a series of representative temperatures of the two heterostructures with rutile (a) and anatase (b) capping on CFO, respectively. Blue and orange curves correspond to the external magnetic field sweeping from 14 T to -14 T and reverse, respectively. Gray curves in (a) and (b) are the *M-H* curve of CFO, with a scale-bar of 200 emu/cm$^3$. $H_c$, coercive field, defined by the sign reversal from -14 T to +14 T.

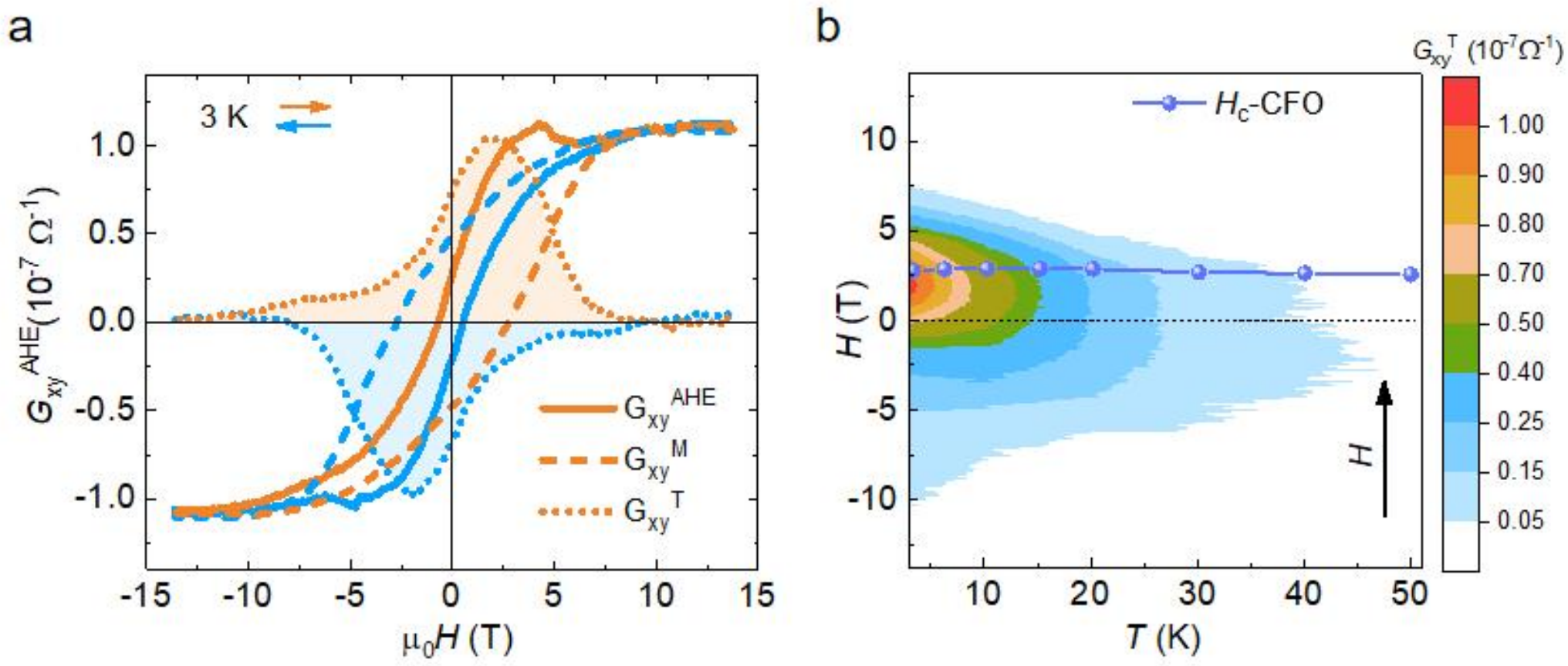

**Figure 4. Analyses of the anomalous Hall conductance curves in the anatase-$IrO_2$/CFO heterostructure. a**, Two components of the anomalous Hall conductance ($G_{xy}^{AHE}$) at 3 K. The measured $G_{xy}^{AHE}$ (solid line) can be divided into two contributions, i.e., the term $G_{xy}^{M}$ derived from the uniform magnetization and that $G_{xy}^{T}$ generated by spin chirality (non-coplanar order). **b**, Contour plot of the $G_{xy}^{T}$ in the $H$-$T$ space. The data were measured with magnetic field sweeping from -14 T to 14 T. The coercive fields ($H_c$) of the CFO with different temperatures are shown.